\documentclass[%
 aip,
 amsmath,amssymb,
 reprint,%
]{revtex4-1}

\usepackage{graphicx}
\usepackage{dcolumn}
\usepackage{bm}

\usepackage[utf8]{inputenc}
\usepackage[T1]{fontenc}
\usepackage{mathptmx}

\draft 

\begin{document}


\title{10 GHz Generation with Ultra-Low Phase Noise via the Transfer Oscillator Technique}



\author{N. V. Nardelli}
 \email{nicholas.nardelli@nist.gov.}
  \affiliation{ 
National Institute of Standards \& Technology, 325 Broadway, Boulder, CO 80305, USA
}%
\affiliation{%
University of Colorado, Boulder, Colorado 80309, USA
}%

\author{T. M. Fortier}%
 \email{tara.fortier@nist.gov.}
\affiliation{ 
National Institute of Standards \& Technology, 325 Broadway, Boulder, CO 80305, USA
}%

\author{M. Pomponio}
 \affiliation{ 
National Institute of Standards \& Technology, 325 Broadway, Boulder, CO 80305, USA
}%
\affiliation{%
University of Colorado, Boulder, Colorado 80309, USA
}%

\author{E. Baumann}
 \affiliation{ 
National Institute of Standards \& Technology, 325 Broadway, Boulder, CO 80305, USA
}%
\affiliation{%
University of Colorado, Boulder, Colorado 80309, USA
}%

\author{C. Nelson}
 \affiliation{ 
National Institute of Standards \& Technology, 325 Broadway, Boulder, CO 80305, USA
}%

\author{T. R. Schibli}
\affiliation{%
University of Colorado, Boulder, Colorado 80309, USA
}%

\author{A. Hati}
 \affiliation{ 
National Institute of Standards \& Technology, 325 Broadway, Boulder, CO 80305, USA
}%


\date{\today}

\begin{abstract}
 Coherent frequency division of high-stability optical sources permits the extraction of microwave signals with ultra-low phase noise, enabling their application to systems with stringent timing precision. To date, the highest performance systems have required tight phase stabilization of laboratory grade optical frequency combs to Fabry-Perot optical reference cavities for faithful optical-to-microwave frequency division. This requirement limits the technology to highly-controlled laboratory environments. Here, we employ a transfer oscillator technique, which employs digital and RF analog electronics to coherently suppress additive optical frequency comb noise. This relaxes the stabilization requirements and allows for the extraction of multiple independent microwave outputs from a single comb, while at the same time, permitting low-noise microwave generation from combs with higher noise profiles. Using this method we transferred the phase stability of two high-Finesse optical sources at 1157 nm and 1070 nm to two independent 10 GHz signals using a single frequency comb. We demonstrated absolute phase noise below -106 dBc/Hz at 1-Hz from carrier with corresponding 1 second fractional frequency instability below $2\times10^{-15}$. Finally, the latter phase noise levels were attainable for comb linewidths broadened up to 2 MHz, demonstrating the potential for out-of lab use with low SWaP lasers. 
\end{abstract}

\pacs{}

\maketitle 

\section{Introduction}

High-stability microwave sources are ubiquitous in modern technology, underpinning communication, computing, and RADAR and sensing systems. Most of the latter systems rely on room-temperature crystal oscillators as local oscillators and frequency references. These commercially available electronic sources, however, exhibit phase noise that is unlikely to support next-generation systems that will require improved resolution and signal-to-noise ratios for higher density communications, micro-Doppler radar, and quantum sensing. More specifically, in Doppler radar, lower close-to-carrier phase noise would increase sensitivity to slow-moving objects, or to objects with low cross section \cite{Ye2000, Fortier2011, Xie2017}. This technique may also aid in the advancement of near-Earth asteroid mapping for both collision warning systems and for the detection of minerals for use on earth or in space \cite{asteroids2015}. In wireless telecommunications, higher spectral purity would also permit lower error rates for phase-encoded signals, and thus higher throughput \cite{Koenig2013}. Finally, sources with lower timing jitter could increase the resolution in analog-to-digital converters \cite{Valley2007}.

Photonically generated microwave signals exploit the high Quality-factors of ultra-stable optical references \cite{Schioppo2016,Young1999} to permit extraction of frequency-divided signals with ultra-low phase noise and high stability. To date, X-band signals near 10 GHz with the highest spectral purity have been achieved via optical frequency division (OFD) using self-referenced optical frequency combs (OFC) \cite{Zhang2010,Haboucha2011,Fortier2013,Xie2017,Kalubovilage2020,Nakamura2020}. Optical frequency combs, based on mode-locked femtosecond lasers, enable a phase-coherent link between the optical and microwave domains. When an OFC is phase stabilized to a ultra-narrow linewidth cavity-stabilized laser, designed to probe a long-lifetime atomic clock transition, it can permit extraction of a 10 GHz microwave signal with phase noise below -100 dBc/Hz at 1 Hz offset from carrier and with a signal-to-noise ratio close to 180 dB in a 1-Hz bandwidth \cite{Fortier2013}. Consequently, when compared to other room-temperature oscillators, OFD allows for greater than 50 dB improvement in close-to-carrier phase noise sidebands. 

Although OFD via phase-locked OFC (PL-OFD) has demonstrated record performance in the domain of room-temperature microwave generation, it suffers two drawbacks. Firstly, it requires tight phase locking of the OFC to an optical reference. This can result in additive instability, and loss of timing due to cycle slips, if noise on the modelocked laser cannot be adequately suppressed by feedback actuators with limited bandwidth. Secondly, tight phase locking of an OFC generally only permits the derivation of low-noise microwave signals from a single optical reference. Consequently, each optical reference requires a corresponding OFC for division to the microwave regime, which may hinder technological implementation due to the increased power consumption and size.

The transfer oscillator (TO) technique \cite{Telle2002} for OFD (TO-OFD) provides solutions to both of the aforementioned limitations. Rather than relying on physical actuators for phase-locking, the transfer oscillator technique employs electronics to add, subtract and divide RF signals in order to remove OFC phase noise from the divided-down optical reference. Due to the absence of physical actuators, the bandwidth of the TO electronics can be several megahertz, decoupling the optical-to-microwave division from the dynamics of the OFC and increasing the robustness of the generated microwave signals. Additionally, the TO technique can simultaneously divide multiple independent optical references with a single OFC and enable the generation of multiple independent low-noise microwave signals. These advantages help support the development of low-noise microwave flywheels derived from optical references, which are needed for realization of optical atomic time for future redefinition of the SI second \cite{Yao2019}. Furthermore, TO-OFD may help to improve multi-frequency radar, which can provide high-resolution images for resource exploration and military ventures \cite{Pfitzenmaier2019}. The TO technique may also help facilitate low-noise microwave generation from chip-scale OFCs that are difficult to stabilize \cite{Lucas2020} or from highly robust polarization-maintaining linear fiber lasers \cite{Baumann2009,Sinclair2014,Feng2015,Cingoz2011} that exhibit higher intrinsic noise. 

In this work, we employ a free-space Er/Yb:glass optical frequency comb as a transfer oscillator to derive low phase noise 10 GHz microwave signals from optical reference cavities at 1157 nm and 1070 nm that serve as local oscillators for the Yb-lattice \cite{McGrew2018} and single-ion Al \cite{Brewer2019} clocks at NIST. Although the optical clocks were not operational for this work, the free-running optical references provided light with high-spectral purity and low drift ($\ll$ 1 Hz/s). From the latter optical references we extracted two independent 10 GHz signals, using a single OFC, with absolute phase noise below -106 dBc/Hz at 1-Hz offset from carrier, and 1-s fractional frequency instabilities below $2\times10^{-15}$. To the best of our knowledge, the results we obtained represent the lowest close-to-carrier phase noise reported to date using the transfer oscillator technique, providing suppression of OFC noise in excess of 100 dB below a 1 Hz offset. It is also the first demonstration of the synthesis of multiple microwave signals based on independent reference cavities with a single OFC.

\begin{figure*}
\includegraphics[width=17cm]{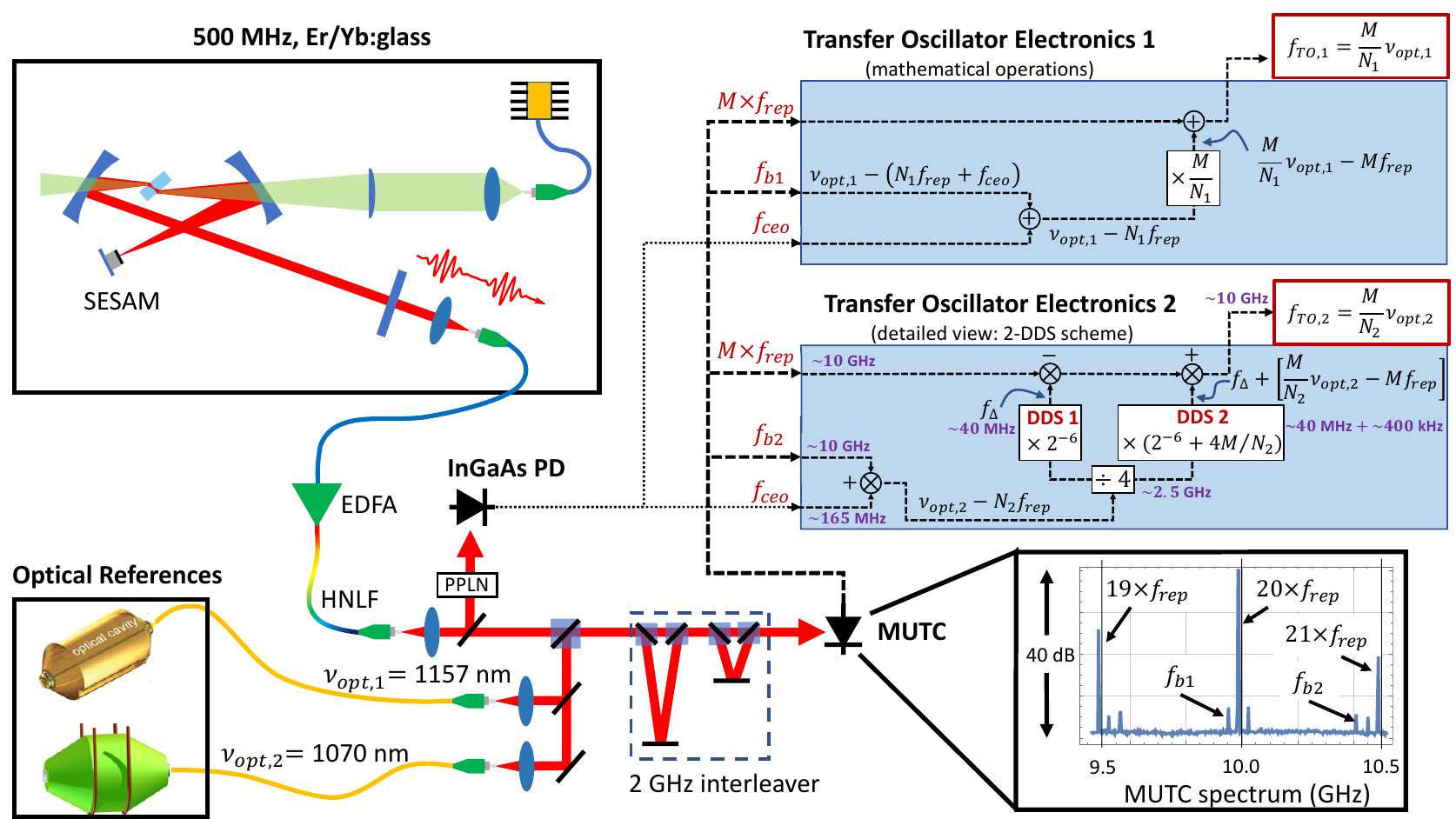}
\caption{Transfer oscillator setup with a 500 MHz repetition rate Er/Yb:glass OFC. Optical pulses from the Er/Yb:glass oscillator are fiber-coupled and amplified by an EDFA before being sent through an HNLF for supercontinuum generation. The spectrum is split in two paths, one for carrier-envelope offset frequency detection, and one for detection of heterodyne beat signals between the OFC and light from two cavity-stabilized optical references at 1157 nm and 1070 nm. An optical interleaver multiplies the repetition frequency to 2 GHz, which is detected on an MUTC photodiode whose spectrum near 10 GHz is shown in the bottom-right inset (RBW = 300 kHz). The optical beats, $f_\text{b1}$ and $f_\text{b2}$, and the 10 GHz harmonic of the 2 GHz pulses are sent to two TO circuits along with $f_\text{ceo}$. Each circuit generates a 10 GHz output, $f_\text{TO,i}$, that is derived from one of the two optical references by adding and dividing microwave signals, shown in blue boxes in upper right. Although the two TO electronics boards are identical except for bandpass filters to isolate $f_{bi}$, the top board in the figure is used to represent the transfer oscillator mathematical operations and the bottom board shows the 2-DDS scheme as well as the approximate RF frequencies at each stage. The following integer constants are used: $M = 20$, $N_1 = $ 518,907 and $N_2 = $ 561,211.}
\label{fig:setup}
\end{figure*}

\section{Experimental Setup}

Photonics-based microwave generation can be separated into three distinct experimental stages: 1) the optical reference that defines a lower limit to the system stability and phase noise, 2) the phase-coherent division performed by the optical frequency comb that enables faithful transfer of the optical reference stability and phase noise to the microwave domain and, 3) the electronic detection and synthesis, which includes photodetection as a means to demodulate the optical signals, and here, a digital transfer oscillator technique that enables electronic removal of OFC noise from the photodetected X-band microwave signals. In the sections below, we describe the details of constituent components in our low-noise microwave generation setup and their impact on the derived 10 GHz microwave signals.

\subsection{Low-noise optical references}

In this work, we employed optical references derived from lasers stabilized to state-of-the-art optical Fabry-Perot cavities \cite{Schioppo2016,Young1999}. As mentioned previously, the optical reference sets the ultimate limit to the noise performance in photonic microwave generation via optical frequency division with an OFC. To minimize instabilities resulting from environmental effects, the cavities are housed within several layers of thermal and acoustic shielding and employ both passive and active vibration control to isolate the cavity length from changes due to acceleration. The room-temperature Fabry-Perot cavities ($\sim$ 30 cm long) act as length references, that at the time of the writing of this manuscript, can support fractional length instabilities near 1 part in $10^{16}$. These cavities act as optical frequency filters when a laser is stabilized to one of their longitudinal optical modes, transferring the cavity length stability to frequency stability of the laser light. Currently, the uncertainty in the cavity length, for timescales greater than 1 second, is limited by thermally induced Brownian noise in the mirror coatings and by unidirectional length changes in the cavity due to settling of the ultra-low expansion glass that comprises the spacer. As such, these optical references can support phase noise as low as -120 dBc/Hz at a 1 Hz offset on a 10 GHz carrier and greater than 200 dB of signal-to-noise ratio in a 1 Hz bandwidth.

Shorter and smaller cavities, more suitable for portable optical atomic clocks and that can support out-of-lab use, are being actively developed \cite{DavilaRodriguez2017, Kelleher2021}. The negative impact of the shorter cavity lengths on the fractional instability can be balanced by optical techniques to improve the thermal noise and by employing crystalline mirror coatings with higher mechanical stiffness, permitting fractional instabilities near 1 part in 10$^{15}$ for cavity lengths of approximately 25 mm \cite{Cole2013}.  

It has also been demonstrated that a free-running OFC can serve as its own optical reference cavity while simultaneously performing the optical-to-microwave division \cite{Kalubovilage2020}. By employing a high-stability monolithic modelocked laser, oscillator length fluctuations are minimized while simultaneously generating a coherent spectrum with which to link optical frequencies to microwave frequencies. While less complex, this solution only permits low phase noise at offset frequencies > 10 kHz. Drift in the modelocked laser cavity yields significantly higher noise than is possible when using Fabry-Perot optical reference cavities.

\subsection{Optical-to-microwave division: Er/Yb:glass OFC}

Optical frequency combs, based on modelocked laser spectra, permit phase coherent transfer of signals between the optical and microwave domains due to the fixed phase relationship between the resonant optical modes. As a result of this phase relationship, the optical spectra of OFCs consist of hundreds of thousands to millions of equidistant frequency modes where each individual mode of the OFC optical spectrum can be described by the following comb equation,
\begin{equation}
\nu_N = f_\text{ceo} + N f_\text{rep},
\label{eq:one}
\end{equation}
where $f_\text{ceo}$ (carrier-envelope offset frequency) is a frequency offset common to each optical mode, $N$ is an integer representing the optical mode number and $f_\text{rep}$ (repetition frequency) is the optical mode spacing. We denote optical frequencies with the "$\nu$" symbol and RF frequencies with the "$f$" symbol. Interference of the OFC light with single frequency light from an optical reference yields a heterodyne beat signal, $f_b = \nu_\text{opt} - \nu_N$, between the optical frequency $\nu_\text{opt}$ and a nearby OFC mode $\nu_N$. 

By rewriting Equation \ref{eq:one} as $(\nu_N-f_\text{ceo})/N = f_\text{rep}$, it is evident that frequency deviations on $f_\text{rep}$ are $N$ times smaller than those of the optical mode, $\nu_N$. As a result, by referencing an OFC to a high-stability and low-noise  optical reference, $\nu_\text{opt}$, this frequency relationship between the optical and mode spacing of the OFC can be exploited to derive microwave signals with high spectral purity. Electronic signals derived in this manner also preserve the stability of the optical reference, such that, ideally $\sigma = \frac{\delta \nu}{\nu} =\frac{\delta f}{f}$. Additionally, photonically generated microwave signals permit a reduction in the phase noise of the optical reference by $(N/M)^2$ when photodetecting the $M$th harmonic of the OFC repetition rate, $f_\text{rep}$. When dividing the 1157 nm (259 THz) and 1070 nm (280 THz) optical reference to 10 GHz, this results in a reduction in phase noise by 88.3 dB and 89.0 dB, respectively.

The optical frequency comb used in our measurements employs a home-built 1550 nm, 500 MHz repetition rate, free-space modelocked laser based on Er/Yb co-doped phosphate glass \cite{Lesko2020}. Passive mode-locking, with an output power around 70 mW, is achieved in the laser via a semiconductor saturable absorption mirror (SESAM). The laser and its supporting optics are illustrated in Figure \ref{fig:setup}.

The 500 MHz optical pulse train generated by the mode-locked laser is fiber-coupled and amplified in an Erbium doped fiber amplifier (EDFA) and then launched into a $\sim 1$ m highly nonlinear fiber (HNLF) to provide nonlinear frequency conversion from the input pulse at 1550 nm (bandwidth $\sim 13$ nm) to an optical octave of bandwidth (1 $\mu$m to 2 $\mu$m). The optical octave is coupled into free space, at which point a part of the spectrum is split off with a dichroic mirror and sent to a periodically-poled lithium niobate (PPLN) waveguide to permit self-referenced detection of $f_\text{ceo}$ via $f$-to-$2f$ conversion \cite{Telle1999, Reichert1999}. The remainder of the light is used to derive optical beat signals against the two high stability optical references at 1070 nm and 1157 nm.

The combined light from the OFC and optical references is sent through a two-stage pulse interleaver and focused onto a highly linear Modified Uni-Traveling Carrier (MUTC) photodetector ($\sim 12.5$ GHz bandwidth, 50 $\mu$m active area diameter, -16 V bias) \cite{Li2011,Li2010}. The pulse interleaver effectively multiplies the pulse repetition rate from 500 MHz to 2 GHz \cite{Haboucha2011}, increasing the strength of the recovered 10 GHz microwave tone, which consequently improves the thermally detected noise floor by mitigating photodetector nonlinearities that result at high optical pulse energies. Since the optical beat signals due to interference between the optical references and optical modes of the OFC are detected on the same MUTC photodiode, the RF output is split with a 1-to-4 power splitter and the optical beats and 10 GHz carrier are isolated from one another with narrow electrical bandpass filters. 

Since TO-OFD permits the extraction of microwave signals via removal of the additive OFC noise, the technique only requires control of the modelocked laser oscillator length such that the optical beat does not drift outside a range bounded by $f_\text{rep}/2$. In this experiment, the narrow-bandpass filters, mentioned above (bandwidth $\approx 5$ MHz), are required to isolate the optical beats from nearby signals. To keep this beat centered on the narrow electronic filter, we employ a loose frequency lock (bandwith < 10 Hz) by monitoring $f_\text{rep}$ and feeding back on the OFC cavity length. This maintains the optical beat close to the center of its bandpass filter. The $f_\text{ceo}$ beat signal is uncontrolled as it drifts by only few MHz over the course of multiple days.

\subsection{Microwave generation: detection and transfer oscillator electronics}
    

To generate a 10 GHz signal from the OFC, while simultaneously removing its noise contributions, the TO technique requires the detection of 1) the $M$th harmonic of the OFC repetition frequency, $M f_\text{rep}$, near 10 GHz, 2) the carrier-envelope offset frequency, $f_\text{ceo} < 250$ MHz, and 3) the RF beat between each optical reference (subscript $i$) and the $N_\text{i}$th OFC mode, $f_\text{b,i} = \nu_\text{opt,i} - \nu_{N_\text{i}}$ also coincidentally near 10 GHz. The harmonic $M f_\text{rep}$ is chosen such that it is close to the desired output microwave carrier frequency of 10 GHz. 
Here, $f_\text{b,i}$ was chosen to be near 10 GHz due to the availability of X-band electronics.

The three beat signals above are mixed such that
\begin{equation}
f_\text{TO,i} = M f_\text{rep} + \frac{M}{N_\text{i}} \left( f_\text{b,i} + f_\text{ceo} \right) = \frac{M}{N_\text{i}}\nu_\text{opt,i},
\label{eq:two}
\end{equation}
which yields a 10 GHz microwave signal free of OFC noise. The resultant microwave signal is dependent only on the static values of $N_\text{i}$ and $M$ and the optical reference frequency $\nu_\text{opt}$. The electronics to facilitate the above signal processing are displayed pictorially in Figure \ref{fig:setup} and described in detail below.

\textbf{1) Signal detection:} In our demonstration, we select the $M=20$ harmonic of $f_\text{rep}$ to generate a microwave carrier signal near 10 GHz. The high-power, high-linearity MUTC photodiode and the optical interleaver help to generate 1 dBm of microwave power at 10 GHz. The OFC mode numbers $N_\text{i}$ with which the two optical references at 1157 nm and 1070 nm interfere are $N_1=518,907$ and $N_2=561,211$, respectively. From this, two optical heterodyne beat signals, $f_\text{b,1}$ and $f_\text{b,2}$, are derived near 10 GHz, with signal strengths varying from -80 dBm to -65 dBm. A separate InGaAs photodetector (EOT 3000A) is used to detect $f_\text{ceo}$. 

\textbf{2) Removal of \boldsymbol{$f_\text{ceo}$} noise:} Each detected heterodyne optical beat signal is routed to its own TO circuit and sent through a series of high-Q bandpass filters (bandwidth $\approx 5$ MHz) to remove the high-power harmonics of $f_\text{rep}$. The filtered signal is then amplified to 10 dBm. The amplified beat is then used to drive the LO port of an IQ mixer (Marki MMIQ-0520L). The OFC offset frequency and its corresponding noise, ($f_\text{ceo}\approx$ 165 MHz), are removed from $f_\text{b,i}$ using an IQ mixer and hybrid coupler with 30 dB of image rejection. This permits the addition or subtraction of signals without a bandpass filter to suppress the unwanted sideband. The result of this operation is $f_\text{b,i} + f_\text{ceo} = \nu_\text{opt,i} - N_\text{i} f_\text{rep}$ where $f_\text{b,i}$ has been expressed using Equation 
\ref{eq:one}. The removal of $f_\text{ceo}$ has been previously demonstrated in photonic microwave generation \cite{Millo2009} for the purpose of reducing the number of feedback loops.

\textbf{3) Removal of optical noise on \boldsymbol{$f_\text{rep}$}:} The goal of this step is to transform the $N_\text{i} f_\text{rep}$ on the right-hand side in the equation in step 2 above to $M f_\text{rep}$ such that the outcome is $\frac{M}{N_\text{i}}[f_\text{b,i} + f_\text{ceo}] = \frac{M}{N_\text{i}} \nu_\text{opt,i} - Mf_\text{rep}$. With $f_\text{ceo}$ removed, this latter signal is achieved by dividing the optical beat ($\approx 10.165$ GHz) by $N_1/M \sim 25,900$ or $N_2/M \sim 28,000$ to scale the optical OFC noise on the $N_\text{i}$th mode to $M f_\text{rep}$. The $f_\text{ceo}$-free optical beat acts as the clock source for a Direct Digital Synthesizer (DDS, Analog Devices 9914), which divides the input by a factor of $N_\text{i}/M$, while also exhibiting low residual phase noise \cite{Fortier2016}. Because the maximum input clock frequency of the AD9914 is 3.5 GHz, we first digitally divided the 10.165 GHz amplified optical beat by 4. We adjusted for this pre-division by reducing the DDS division by a factor of 4.

To reach the final transfer oscillator output,  $f_\text{TO,i} = \frac{M}{N_\text{i}} \nu_\text{opt,i}$, the additional $Mf_\text{rep}$ must be removed from the right-hand side such that $\frac{M}{N_\text{i}}[f_\text{b,i} + f_\text{ceo}] + Mf_\text{rep} = \frac{M}{N_\text{i}} \nu_\text{opt,i}$. The DDS-divided optical beat (near 400 kHz) is mixed with $M f_\text{rep}$ directly out of the photodiode, using an IQ mixer, to remove repetition rate noise contributed by the OFC. To drive the LO port of the mixer, the $Mf_\text{rep}$ signal is amplified to 10 dBm with low phase noise amplifiers (Custom MMIC CMD245), which help preserve the ultra-low phase noise of the microwave carrier directly out of the OFC. This last step removes the multiplicative noise of $f_\text{rep}$, permitting the derivation of a microwave signal that only contains the frequency and phase noise of the optical reference. Additional optical references may be divided simultaneously with the OFC, each resulting in a corresponding microwave signal.

\textbf{Phase noise spur suppression with two DDSs:} The DDS output frequency $f_\text{DDS} = \frac{M}{N_\text{i}}[f_\text{b,i} + f_\text{ceo}]$ is close to 400 kHz, which, when mixed with $Mf_\text{rep}$, results in spurs that are harmonics of 400 kHz and that pose a challenge to filtering. To minimize these harmonics, two DDSs are employed to perform a dual heterodyne frequency translation \cite{Brochard2018}. DDS 1 synthesizes a frequency $f_\Delta$ and DDS 2 synthesizes $f_\Delta + f_\text{DDS}$. Here, $f_\Delta$ is a frequency large enough to be filtered from the 10 GHz signal. We chose $f_\Delta \approx 40$ MHz, which corresponds to a division factor of $2^6$, i.e., 2500 MHz / $2^6 \approx$ 40 MHz. 

The DDS 1 signal is subtracted from $M f_\text{rep}$ using an IQ mixer. The resulting signal ($Mf_\text{rep} - f_\Delta$) is then added using a second IQ mixer to the DDS 2 signal. The result, ($Mf_\text{rep} + f_\text{DDS}$), yields the same frequency as the 1-DDS scheme, but spurs due to mixing products are pushed to harmonics of $f_\Delta \approx 40$ MHz and its multiples that are more easily filtered from the 10 GHz tone. The 2-DDS scheme is shown pictorially in Figure \ref{fig:setup}.

Imperfect spur suppression is mitigated by using a narrow-band filter (bandwidth $\approx 5$ MHz) between the aforementioned IQ mixers.  While filtering reduces the 400 kHz phase noise spurs by nearly 40 dB, the filter introduces a significant group delay to the 10 GHz carrier signal. To ensure the highest bandwidth suppression of OFC noise, a delay line of several meters was added between DDS 2 and the second IQ mixer to compensate for the group delay of the filter.

\section{Results}

\begin{figure}
\includegraphics[width=8.5cm]{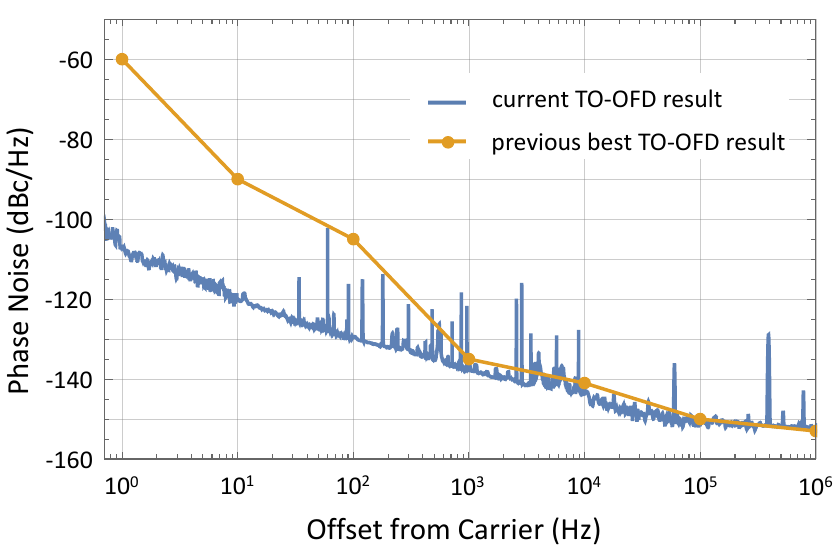}
\caption{Comparison between the current TO result (blue trace), the previous lowest phase noise TO result (connected orange dots) \cite{Brochard2018}.}
\label{fig:comparison}
\end{figure}

In this Section we discuss characterization of the phase noise and stability of the 10 GHz output from our TO-OFD scheme. We specifically assess the difference in performance between OFD derived via TO-OFD and tight phase locking (PL-OFD), and characterize the limiting sources of noise in our TO scheme. Lastly, we demonstrate noise suppression for optical linewidths as large as 2 MHz. 

Figure \ref{fig:comparison} shows how the phase noise spectrum of the current demonstration compares to that of the previous best TO demonstration \cite{Brochard2018}. We show more than 40 dB of phase noise reduction at 1-Hz offset from carrier and a similar performance at high frequencies.

\subsection{10 GHz characterization}

\begin{figure}
\includegraphics[width=8.5cm]{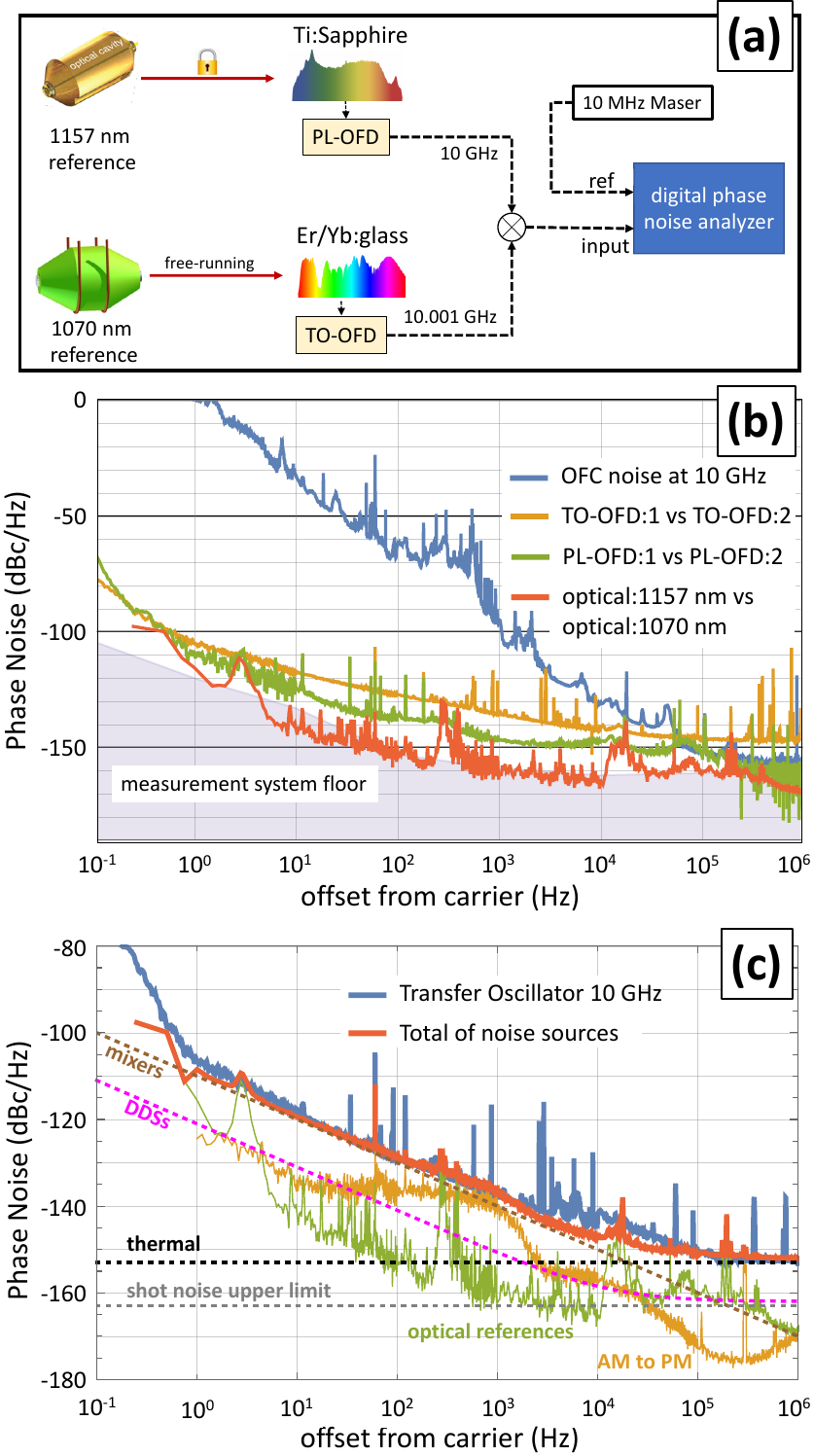}
\caption{a) Transfer oscillator microwave characterization setup. b) Phase noise comparison between $Mf_\text{rep}$ generated by the Er/Yb:glass OFC when unstabilized (blue) and when stabilized to the 1070 nm optical reference (green) and $f_\text{TO,1}$ vs $f_\text{TO,2}$ when the OFC is unstabilized (orange). The absolute 10 GHz limit (red) is set by the relative phase noise between 1157 nm and 1070 nm optical references. c) Phase noise comparison between $f_\text{TO,2}$ derived from the 1070 nm reference (blue) and the total of all calculated noise sources (red). Constituent noise sources are plotted as dashed lines and thin traces.}
\label{fig:phase-noise}
\end{figure}

The 10 GHz microwaves generated by the transfer oscillator technique are compared against a reference 10 GHz signal derived from a Ti:Sapphire OFC \cite{Fortier2006, Fortier2011} (see Figure \ref{fig:phase-noise}a) that is phase-stabilized to the 1157 nm optical frequency reference. A 10 GHz tone, which was detected using an MUTC photodiode, was filtered and sent over approximately 3 meters of microwave cable (Maury Microwave SB-SMAN-MM) to the Er/Yb:glass OFC and TO electronics. The frequency difference ( $\approx$ 1 MHz) between the two 10 GHz signals was measured using a Symmetricom 5125a digital phase noise analyzer. A 10 MHz H-maser signal served as the phase noise reference and contributed flicker phase noise near -120 dBc/f and a white phase noise floor near -160 dBc/Hz. We used this method to characterize the phase noise of each of the two $f_\text{TO,i}$ signals, derived from the 1157 nm and 1070 nm references, separately, and also the phase noise of $Mf_\text{rep}$ generated via PL-OFD when the Er/Yb:glass OFC was phase-locked to one of the optical references.

As mentioned previously, TO-OFD allows for independent microwave signals to be derived from multiple optical references using a single OFC. We generated two distinct $f_\text{TO,i} \approx 10$ GHz microwave signals derived from the 1157 nm and 1070 nm cavity-stabilized lasers using two TO channels on the Er/Yb:glass OFC. The phase noise from this comparison is shown as the orange curve in Figure \ref{fig:phase-noise}b. Also seen in Figure \ref{fig:phase-noise}b, is the free running noise on $Mf_\text{rep}$. Using the TO technique, we observed suppression of unstabilized OFC noise of greater than 100 dB close to the 10 GHz carrier. Correspondingly, we observe fractional frequency instability suppression of the unstabilized carrier from more than $10^{-10}$ to less than $10^{-15}$ after 1 second of averaging.

The measurements in Figure \ref{fig:phase-noise}b all constitute absolute phase noise comparisons between microwaves generated from the 1157 nm and 1070 nm cavity-stabilized laser references. Microwaves produced via TO-OFD exhibited a 1 Hz phase noise level of approximately -106 dBc/Hz, which we observed to be limited by the relative drift between the two optical references. This can be deduced from the relative optical phase noise of the 1157 nm and 1070 nm references, scaled to 10 GHz by a factor of $-20 ~\log (280 \text{ THz}/10 \text{ GHz}) \approx -89$ dB. This measurement (red curve in Figure \ref{fig:phase-noise}b) represents the lower limit of 10 GHz phase noise that could be derived from these optical references.

Additionally, we compared the phase noise of the 10 GHz signal derived via PL-OFD when the Er/Yb:glass OFC was phase-stabilized to the 1070 nm optical reference against the 10 GHz signal generated via PL-OFD of the Ti:Sapphire OFC, phase-stabilized to the 1157 nm optical reference. The resultant phase noise (green trace in Figure \ref{fig:phase-noise}b) demonstrates the contribution of the two OFCs and  photodetection. 
From the comparison between green and orange traces, we observe that the 10 GHz electronic components in the TO technique (mixers, amplifiers, DDSs) limit the phase noise for frequency offsets 20 Hz and above.


\subsection{Phase noise contributions}
Here we discuss the noise contributions to TO-OFD from its various components detailed in Section II. A summary of these contributions is enumerated and plotted in Figure \ref{fig:phase-noise}c. Also shown is the quadrature sum of the individual noise sources (red trace), assuming that they are uncorrelated. Evident in Figure \ref{fig:phase-noise}c, these sources account for most of the phase noise observed on the $f_\text{TO,i}$ signal (blue trace).

For offset frequencies less than 10 kHz, we observe that noise from the IQ mixers, which contribute flicker phase noise near -110 dBc/f, ultimately limit $f_\text{TO,i}$ phase noise. At high frequencies, the phase noise is limited by thermal noise, the dominant contributors of which are the two IQ mixers and three low phase noise amplifiers that combine the $M f_\text{rep}$ and the 40 MHz DDS outputs. The noise figures of the individual components, 10 dB for each mixer and 3 dB for each low phase noise amplifier, act in quadrature on the -7 dBm $M f_\text{rep}$ signal to yield a thermal limited noise floor of -153 dBc/Hz. We estimate that photodetector shot noise on $M f_\text{rep}$, which takes into account photocurrent contributions from the the single frequency 1157 nm and 1070 nm light (1 mA) and the pulsed light from the OFC (7 mA), is less than -163 dBc/Hz \cite{Quinlan2013}.

The DDS pairs employed in the TO circuits exhibit flicker (-121 dBc/f) and quantization noise (-162 dBc/Hz) on the $f_\Delta \approx 40$ MHz signal. One drawback of employing DDSs is that they contribute phase truncation spurs, with frequencies that depend on the division factor $M/N_\text{i}$ and the frequency offset $f_\Delta$. These spurs can be minimized by setting the DDS input/output ratio to be an exact multiple of 2, $f_\text{out} = 2^{-n} f_\text{in}$ \cite{Gentile1}, where $n$ is a positive integer. It was not possible to implement this strategy in both DDSs because the DDS outputs are necessarily different. While not employed here, for applications that do not tolerate spurs, "bright" spurs can be traded for a slightly raised white noise floor by dithering the last bit in the DAC \cite{Gentile2}.
Finally, despite employing two DDSs to minimize the 400 kHz sidebands that result from $[f_\text{b,i} + f_\text{ceo}] \times M/N_\text{i} \approx 400$ kHz, imperfect mixing and filtering results in a residual spur at 400 kHz.

Due to the nonlinear elements in the TO setup (i.e., the photodiode, amplifiers, 10 GHz mixers) it is important to consider the effect of amplitude noise to phase noise conversion. To this end, we characterized the total amplitude to phase noise conversion for our setup, including the total contributions of the 10 GHz photodetector, TO electronic circuit and 10 GHz phase noise comparison electronics. The characterization of the AM-to-PM conversion coefficient is achieved by adding white amplitude noise to the laser by modulating one of the EDFA 980 nm pump diodes and measuring the RIN with a slow photodiode (Thorlabs PDA20CS). With the added RIN we remeasured the $f_\text{TO,i}$ phase noise and discerned a single-sided AM-to-PM conversion of approximately -32 dB. As seen via the orange trace in Figure \ref{fig:phase-noise}c, using this conversion coefficient, we estimate the 10 GHz phase noise converted from RIN on the light incident on the MUTC photodetector (without added noise). Much of the RIN we observed in this measurement originated from the output of the semiconductor optical amplifier used to amplify the 1157 nm reference light. Amplification was not also required for the 1070 nm reference light since the OFC generated sufficient optical power at this wavelength to generate a high-SNR beat.

Finally, due to the relatively high flicker contribution from the IQ mixers, the combined phase noise of the optical references is only observed to impact the phase noise spectrum of $f_\text{TO,i}$ at low frequency offsets (less than 1 Hz). This noise is visible as a slope change in the absolute phase noise measurements that compare $f_\text{TO,1}$ and $f_\text{TO,2}$ derived from the 1157 nm and 1070 nm references. This thermal noise limit can be improved by operating the cavities at cryogenic temperatures, or by employing high-quality factor mirrors with crystalline coatings \cite{Cole2013}.

\subsection{Frequency noise suppression}

\begin{figure}
\includegraphics[width=8.5cm]{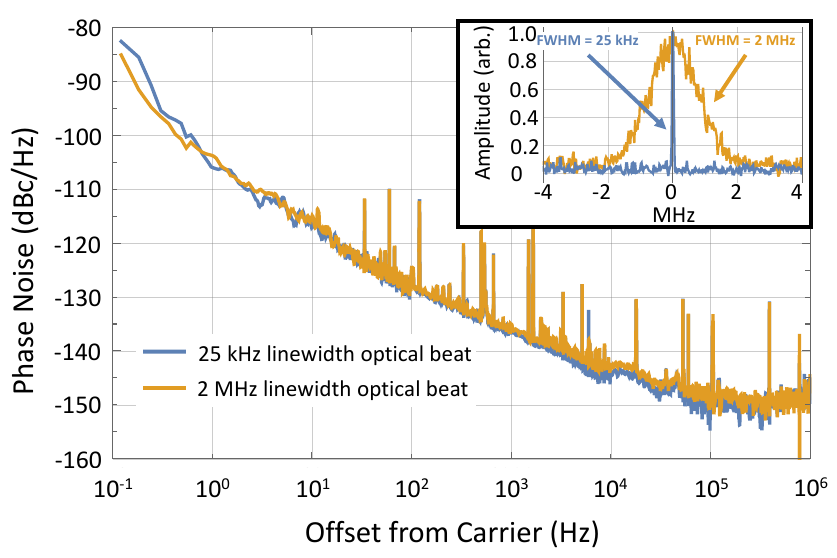}
\caption{Demonstration of noise suppression. Transfer oscillator microwave phase noise without added OFC frequency noise (blue) and with added OFC frequency noise (orange). Inset: the optical beat between a cavity stabilized laser and the comb, without and with added noise (blue and orange respectively). Without added noise the optical linewidth is 25 kHz and with added noise the optical linewidth is 2 MHz.}
\label{fig:noise-suppression}
\end{figure}

Due to the extremely high bandwidth of the digital and RF electronics, in principle, TO-OFD should permit near perfect cancellation of additive OFC noise for optical linewidths in excess of 1 MHz. As mentioned previously, noise suppression for microwave generation using the TO technique has direct application to field-deployable OFCs. In an attempt to quantify the cancellation, we increased the frequency noise on the OFC such that optical mode linewidths increased from 25 kHz to 2 MHz (near 1157 nm) by modulating an OFC laser mirror bonded to a fast PZT (PI Ceramic PL055.31) with white frequency noise. As seen in Figure \ref{fig:noise-suppression}, we compare $f_\text{TO,i}$ with and without added frequency noise. We observe nearly identical phase noise spectra despite the difference in OFC optical linewidths. The narrow ($\approx 5$ MHz) bandwidth of 10 GHz filters used to isolate the optical beat signals, limit the suppression bandwidth in our TO scheme. Ideally, latency in the TO electronics, including the DDS, will set the noise suppression bandwidth. For the AD9914 DDS used here, the latency of a 2.5 GHz clock is approximately 150 ns. 

Although TO-OFD demonstrates excellent frequency and phase noise suppression, it is unable to suppress noise resulting from AM-to-PM conversion. This may be dealt with by reducing the amplitude noise of the laser, for example, with active RIN suppression, or by operating the photodiode, amplifiers and mixers under conditions where AM-to-PM conversion is naturally suppressed \cite{Xie2017,Taylor2011,Baynes2015}.

\subsection{Long-term behavior}

\begin{figure}
\includegraphics[width=8.5cm]{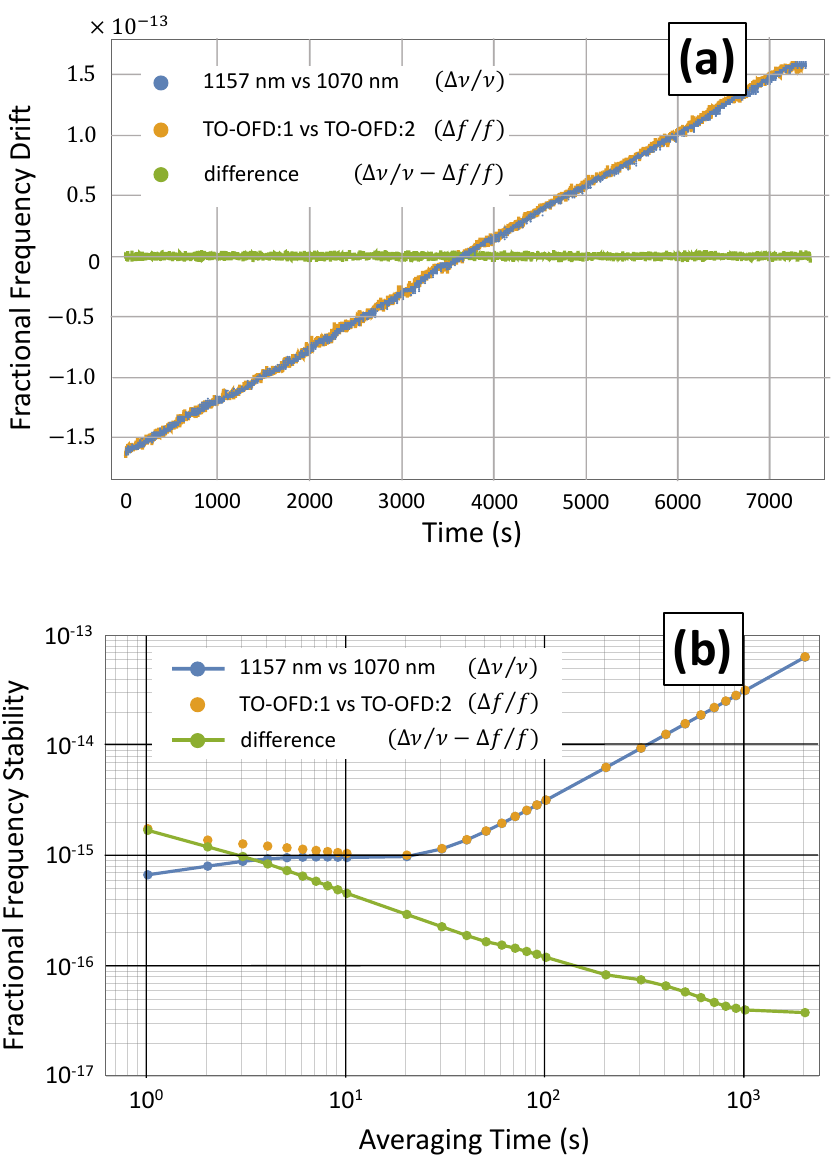}
\caption{a) Time record and b) Allan deviation of the optical beat between the 1157 nm and 1070 nm optical references (blue points) and the microwave beat (orange points) between $f_\text{TO,1}$ and $f_\text{TO,2}$ derived from the same optical references. The microwave beat is normalized by 10 GHz and the optical beat is normalized by 280 THz in a). The difference (green points) between the blue and orange points is shown in a) and its Allan deviation is shown in b).}
\label{fig:counting}
\end{figure}

While phase noise is used to evaluate the short term performance of the $f_\text{TO,i}$ microwave output (timescales less than 1 s), we are also interested in assessing the long-term performance of the microwaves generated via TO-OFD (timescales up to 10,000 s). More specifically, we want to determine the fidelity with which the 10 GHz output reproduces the fractional frequency stability and variations of the the optical references. To measure this, we simultaneously tracked the relative frequency drift of the 1157 nm and 1070 nm reference lasers in the optical domain using the Ti:Sapphire OFC, and the relative frequency drift between the two $f_\text{TO,i}$ signals produced using the Er/Yb:glass OFC from the same two optical references. Frequency data was counted using two Agilent 53131a frequency counters, each referenced to a 10 MHz H-maser signal.

As seen in Figure \ref{fig:counting}a, we observed a strong correlation between the  time records of the fractional frequency deviations in the optical signals and microwave signals. The level of correlation is quantified by the residual deviations in the difference frequency. The Allan deviation evaluates the fractional frequency stability of this data as a function of measurement time. As seen in Figure \ref{fig:counting}b, the upper bound to the residual instability contributed by the Er/Yb:glass OFC and TO electronics is less than $2 \times 10^{-15} \tau^{-1/2}$.

\section{Conclusion}

We have demonstrated microwave generation via optical frequency division of two high-stability optical references with absolute phase noise at or below -106 dBc/Hz at a 1 Hz offset from a 10 GHz carrier using the transfer oscillator technique. The latter result corresponds to a 1-second frequency instability of less than $2\times10^{-15}$, which indicates that the TO digital and analog electronics contribute minimal additional instability. This result represents the lowest-phase noise achieved, to date, using TO-OFD and is the first demonstration generating multiple independent optically derived microwave signals from a single OFC. In addition to characterizing and identifying the noise contributions to our transfer oscillator scheme, we have also demonstrated that the contribution of the OFC's noise to the TO signal is negligible for optical linewidths as large as 2 MHz, and OFC phase noise is suppressed by more than 100 dB close to carrier. This has direct application to low-noise microwave generation with more compact OFCs that exhibit high intrinsic noise. Additionally, TO-OFD provides the means to produce a reliable and robust optically derived microwave signal suitable for out-of-the lab applications.

Finally, the transfer oscillator technique provides significant advantages if used for realization of an optical timescale. Timescales derived from an ensemble of optical references permit a large reduction in frequency noise and a slower accumulation of timing uncertainty \cite{Milner2019} than timescales based on microwave sources and references. The low residual noise and the increase of noise suppression bandwidth provided by TO-OFD will increase operational robustness and reduce the incidence of signal phase slips. Additionally, the ability to derive independent microwave signals from more than one optical reference may facilitate the redundancy necessary for deriving phase-continuous and high-stability electronic timing signals.

\section{Acknowledgements}
We thank I. Coddington, S. A. Diddams and F. Quinlan for their comments on this manuscript and J. C. Campbell for the MUTC photodiode.

\section{Disclaimers}

Certain commercial equipment, instruments, or materials (or suppliers, or software, ...) are identified in this paper to foster understanding. Such identification does not imply recommendation or endorsement by the National Institute of Standards and Technology, nor does it imply that the materials or equipment identified are necessarily the best available for the purpose.

\section{Data Availability}
The data that support the findings of this study are available from the corresponding author upon reasonable request.

\section*{References}


%
%


\bibliography{main.bib}

\end{document}